# PRESY: A Context Based Query Reformulation Tool for Information Retrieval on the Web

[1,2]Abdelkrim Bouramoul, [2]Mohamed-Khireddine Kholladi and [3]Bich-Lien Doan
[1]Department of Computer Science, Guelma University, BP 401 Guelma 24000, Algeria
[2]MISC Laboratory, Department of Computer Science, Constantine University,
BP 325, Constantine 25017, Algeria
[3]Department of Computer Science, SUPELEC, 3 Rue Joliot-Curie, 91192 Gif Sur Yvette, France

**Abstract: Problem statement:** The huge number of information on the web as well as the growth of new inexperienced users creates new challenges for information retrieval. It has become increasingly difficult for these users to find relevant documents that satisfy their individual needs. Certainly the current search engines (such as Google, Bing and Yahoo) offer an efficient way to browse the web content. However, the result quality is highly based on uses queries which need to be more precise to find relevant documents. This task still complicated for the majority of inept users who cannot express their needs with significant words in the query. For that reason, we believe that a reformulation of the initial user's query can be a good alternative to improve the information selectivity. This study proposed a novel approach and presents a prototype system called Profile-based Reformulation System (PRESY) for information retrieval on the web. **Approach:** It used an incremental approach to categorize users by constructing a contextual base. The latter was composed of two types of context (static and dynamic) obtained using the users' profiles. The architecture proposed was implemented using .Net environment to perform queries reformulating tests. **Results:** The experiments gave at the end of this article show that the precision of the returned content is effectively improved. The tests were performed with the most popular searching engine (i.e., Google, Bind and Yahoo) selected in particular for their high selectivity. Among the given results, we found that query reformulation improve the first three results by 10.7 and 11.7% of the next seven returned elements. So as we could see the reformulation of users' initial queries improves the pertinence of returned content. **Conclusion/Recommendations:** Therefore, we believed that the exploitation of contextual data based on users' profiles could be a very good way to reformulate user query. This complementary mechanism would be highly improve the quality of information retrieval on the web. In the other side, we believe that more the user's profiles are properly constructed more the returned documents are relevant. Thus, the approach of constructing profiles needs to be deeply studied in order to have more representative elements. Additional data like historical searching and browsing activity of a user can be also combined to improve the query reformulation. This constitutes a part of our perspectives to improve PRESY.

## INTRODUCTION

The Information Retrieval (IR) can be defined as an activity whose purpose is to locate and deliver a set of documents to a user according to his needs. To improve the performance of such systems, the contextual information retrieval has recently emerged as a priority (Allan, 2003); its goal is to place the user at the core of models by making certain elements of context which can influence systems performance more explicit.

On the other hand, users of information retrieval system are not professionals in the documentation (Lin and Wang, 2006). They do not know choosing the right words that best express their information needs, so query reformulation is needed. This reformulation is motivated by the fact that the initial query returns a result that rarely meets the user's need, that means to modify the original user's query by adding significant terms to give back a more relevant result.

In this study we propose a contextual query reformulation system; this reformulation is called contextual because it takes into account the context via the user's profiles to change his initial query. By adding the notion of the context during the

**Corresponding Author:** Abdelkrim Bouramoul, Department of Computer Science, Guelma University, BP 401 Guelma 24000, Algeria

reformulation, we aim to increase the effectiveness of IRS by improving their relevance and allow the consideration of the user's personal characteristics, his interests, his preferences and the historic of his interactions with a system. These elements are capitalized in our system as a static and dynamic context for later use in the contextual reformulation.

**Classic query reformulation approach:** The user is often unable to express his exact need of information. Therefore, among the documents returned, some are more interesting than others. Given the increasing volumes of information bases, to find those that are relevant by using only the initial user's query is an almost impossible task.

The query reformulation is to modify the user's query by adding significant terms. The idea of query refinement is not new; several approaches use different techniques for select terms to be added to the initial query. We distinguish three types of query reformulation approaches and the deference between them lies on the one hand in the source of terms used in the reformulation that may come from results of previous research (relevance feedback) or from an external resource (semantic networks, thesauri or ontology). On the other hand, it lies in the method used for selecting terms to be added to the initial query.

The first type of the approaches is based on a global analysis of the considered collection of documents and the most commonly among them is based on statistical analysis of document corpus (Cui *et al.*, 2002). The objective is to increase the frequency of words appearing together in one document and select the terms with the highest coefficient. The information thus obtained is used to reformulate query automatically by adding terms related to the terms already used in the query. The terms added from the documents give a better adequacy between the need for information and the document collection.

The second type of approaches based on the principle of relevance feedback aims to reformulate the initial query to correspond better to the content of the documents collection. The principle is as follows; the user submits his initial query and the system returns an initial set of documents that the user has to judge (relevant, irrelevant). Knowing the relevance of initially returned documents is used for selecting terms to be added to the initial query. We quote in this category the work of (Winograd, 2001) in which the system offers, based on the first query, a set of documents and according to those viewed by the user, the system updates its terms index in concordance with automatic learning methods.

The last type of the approaches, described in the literature, uses external resources of terms such as thesauri or ontology that contain the vocabulary used in the query enrichment, such approaches use ontology with equivalence and subsumption relations (Navigli and Velardi, 2003) in order to extract the terms to be added to the initial query.

The reformulation approach proposed in this study is based on taking into consideration the user's context using his profile. It offers a double benefit comparing to the above presented approaches. On the one hand and contrary to the first two classes of approaches, it is used directly without a phase of analysis or of learning, on the other hand it is not constrained by the problem of the third class of the approaches that use only equivalence and subsumption relations and do not take advantage of all semantic relations provided by the ontology.

**Context for information retrieval:** The context is not a new notion in computer science: from the sixties, operating systems, language theory and artificial intelligence already exploited this concept. With the emergence of information retrieval systems, the term was rediscovered and placed at the core of the debates without making subject of a consensus, clear and definitive definition. However, analysis of existing definitions in the literature leads to two conclusions:

- "There is no context without context" (Brezillon, 2003). In other words, the context does not exist as such. It is defined or it emerges for a purpose or precise utility
- "The context is a set of information. This set is structured, it is shared, it evolves and serves the interpretation" (Winograd, 2001). The nature of information and interpretations got from it depend on the purpose

In information retrieval, the context is defined as "All cognitive and social factors as well as the user's aims and intentions during a search session", (Belkin *et al.*, 2004). Generally speaking, the context includes elements of various natures that delimit the understanding, the application fields or the possible choice. The most commonly cited elements concern the spatiotemporal data (location, time, date) or specific knowledge in relation to the studied area. But rarely we see the use of elements concerning the emotions, state of mind cultural information (Brezillon, 2003). Thus some elements of context can be difficult to identify because we use them unconsciously, others are out of reach of machines input devices and therefore they are difficult to be implemented in information retrieval systems.

## MATERIALS AND METHODS

**Aspects of profile relevant to context capturing:** A user profile is defined as "A source of knowledge that includes all the acquisition about the user's aspects that may be useful for the system behavior" (Wahlster and Kobsa, 1986). This proposal, although general, corresponds better to our orientation; it highlights three aspects of profile which are exploited as follows:

- Source of knowledge: The user profile may include a variety of information according to the considered task. In information retrieval the content of a user's profile can be summarized in his personal characteristics, his interests and preferences, his competence, his current goal and the historic of his interactions with the system (Belkin *et al.*, 2004). We note that the notion of context, previously presented, is an extension of the user's profile. The context contains complementary information allowing a better adaptation of the profile
- Acquisitions: The content of the user's profile is knowledge to recover. Depending on the system adaptation degree, the user's profile data can be informed by the user, or recovered by selecting a preexisting profile created by experts in the domain, or captured by the information retrieval system during use
- Useful for the system behavior: In information retrieval the role of the user's profile is to allow personalization or adaptation of services to improve the system performance, or to filter the results returned by a search engine

**Classification of profiles and their use in IR:** Different types of profiles are associated with the task of information retrieval. We defined four classes of profile based on grouping criteria, these criteria are the degree of the user's involvement, the moment to use the profile for reformulation, the complexity of information accumulated by the profile and the nature of information that composes the profile.

**According to the user's involvement:** It is to measure the degree of the user's involvement in the process of his profile catching. In (Benammar *et al.*, 2002), the author distinguishes two types of profiles management:

- Indirect: It's the case where the profile management is transparent to the user, in the sense that the user does not intervene in the management of its profiles

- Direct: In contrast, in the direct management of profiles, the user must intervene in all stages of the research process to manage his profiles

**According to the reformulation moment:** Here we look at the moment to use the profile in information retrieval system; two possibilities are to retain:

- Pre-research: A profile can be used in a pre-research stage to help the user to formulate or reformulate his information need. It can be for example to refine the expression of a query proposed by the user according to his profile
- Post-research: A profile can also be used in a post-research stage to filter the search results

**According to the complexity:** In this category, we focus on the degree of the complexity of the information contained in the profile. Different formats of this type of profile have been studied in (Korfhage, 1997), the most important ones are:

- Simple: Simple profile is presented as a set of keywords and eventually an associated coefficient to measure the importance of each term in the profile
- Extended: Extended profile includes, in addition to keywords and their coefficient, a set of information that describes the research context

**According to the nature of information:** The last class distinct user's profiles based on the nature of information they contain, the work of (Benammar *et al.*, 2002) exploits the following profiles:

- Identification profile: This component of the profile is used to identify the user through a set of information. It is defined in the first connection to the profiles system and is updated at each increment by creating a querying profile
- Querying profile: It can be likened to a query. It reflects the user's information needs and facilitates the association of the research made by the user to his context

The different classes of profiles that we identified can be used simultaneously in the same system and the usage of a profile type does not mean the isolation of the others. However, a profile based system must define the characteristics of each type of used profiles and the links that connect them. In the rest of this study we present our system and justify the choice of its parameters in terms of profile, we also describe the proposed architecture and the developed prototype.

**Choice of system parameters:** We have previously prepared a comparative study of the elements needed to the development of our system. They are the reformulation, the context and the use of profiles in information retrieval systems. This study has allowed us to categorize separately the characteristics of these elements and identify the limits of each category. We present here the different parameters characterizing our proposal and describe our choices compared to the approaches previously studied.

**In terms of query reformulation:** The reformulation approach that we propose in this study is based on taking user's context into consideration via his profile. The contextual reformulation system proceeds an automatic refinement or extension of the initial query by adding or substituting words. The terms of reformulation are extracted from the user's profile depending on the context of the underway research session. Table 1, locates our approach compared to the approaches previously studied.

**In terms of the profile use:** Our choice was fixed on the use of context for the user's query reformulation. We presented previously the different classes of profiles and the specific aspects of each class. Table 2, presents the parameters characterizing our system in terms of user's profile.

Table 2 shows our choices concerning the user's profiles and therefore the manner in which the context is used to help the contextual query reformulation. These choices are interpreted as follows:

- User involvement: In our system the user intervenes partially to define his profile, so the involvement is direct; the system automatically recovers the supposed relevant information to enrich the user's context for possible future research and provides them to the user, who will validate those he deems truly relevant among all proposals
- Moment of reformulation: In our case it is a pre-research profile. The system reformulates the user's need by refining the expression of his query according to his context
- Degree of complexity: The profile is extended. It includes, in addition to keywords, a set of information describing the research context. This information is stored in a table as an attribute-value pairs, where each pair represents a property of profile
- Nature of Information: We use at the same time and in a complementary manner identification and querying profiles. The first to identify a user using a set of information defined in the first connection to the system and the second is recovered from the historic search of the same user in the subsequent sessions, so its content increases every time the user makes a new search.

**Summary:** We group our choices in terms of profile use for the context modeling into two major classes:

- Static Context: It is defined from an identification profile with an extended degree of complexity, an indirect involvement of the user and a pre-research moment of use
- Dynamic Context: It is defined from a querying identification profile with an extended degree of complexity and an indirect involvement of the user and a Pre-research moment of use:

  Static Context = (Identification profile + Extended +Pre-research + direct)
  Dynamic Context = (Querying profile + Extended +Pre-research + direct)

**System architecture:** We propose a contextual query reformulation system for information retrieval on the web. This reformulation is called contextual because it is based on coupling between the static and the dynamic user's context that are modeled by using the user's profiles to allow the expansion of the initial query. The goal is to return a result more relevant than the one returned by a search without reformulation. Figure 1, shows the general architecture of our system and illustrates how static and dynamic contexts are caught to be used later in the contextual reformulation process.

Table 1: Characteristics and limits of the approaches for query reformulation

| Class of approaches | Characteristics | Limits |
|---|---|---|
| Based on statistical analysis | Calculates the frequency of terms in a document Selects the terms with the highest coefficient | Requires a analysis phase |
| Based on relevance feedback | Performs a first search using initial query terms Relevant documents returned by using the first query are used to refine the search | Requires a learning phase |
| Based on terms of external resources | Uses ontology or thesauri containing the vocabulary used in the enrichment of query | Does not take advantage of all semantic relations provided by ontology |
| Our approach | | |
| Based on static and dynamic context | Uses the context via the user's profile to reformulate automatically the initial query by adding terms extracted from the context of current research | |

Table 2: Choice of system parameters in terms of the use of profile

| Criteria | Our choice | |
|---|---|---|
| User involvement | Direct X | Indirect |
| Moment of reformulation | Pre-research | Post-research X |
| Degree of complexity | Extended X | Simple |
| Nature of information | Querying profile X | Identification profile X |

Before formulating his query, the user must be identified in the system which proceeds to the recovery of its static context that consists of the user's personal characteristics that may influence the research context (age, sex and language), his interests and preferences related to the research task (domains of competence and specialty) and his competence level or his expertise (profession and level of study). This information is stored in the user's context base during the first connection to the system. In case of a user who does not have a profile, the system asks him to complete his preference and the user's context base will be updated in order to use it in any future search sessions.

After having recovered the static context, the user can formulate his query and the system takes charge of his reformulation automatically. It generates the new query by selecting terms related to the context of the current research session. This selection is made from the user's context base. Both types of context (static and dynamic) mutually contribute to the operation of reformulation. Thereafter the system proceeds to an open search on the web by using the reformulated query and calling, according to the user's choice, one of the three search engines that suggests (google, yahoo or bing). The search result is finally returned to the user and it will also be stored in the historic search base in order to be used later in the catching of the dynamic context.

At the end of each search session and based on the historic search base, the system automatically retrieves the information that supposed relevant to enrich the dynamic context of the user (contextual elements). The system suggests this information to the user to validate those he considers truly relevant among all proposals and increases the content of the dynamic context.

**RESULTS**

To prove the applicability of the proposed architecture, we implemented a prototype of a system for contextual query reformulation. PRESY is a .Net application using a Microsoft Access data base which manages the users' profiles. This database contains two tables, the first serves to save the user's preferences (static context) and the contextual elements (dynamic context) and the second table contains the historic of the user's research that is used to catch the dynamic context.

**General description:** The system displays two results. The first without reformulation (Fig. 2, area A) and the second with reformulation (Fig. 2, area B).

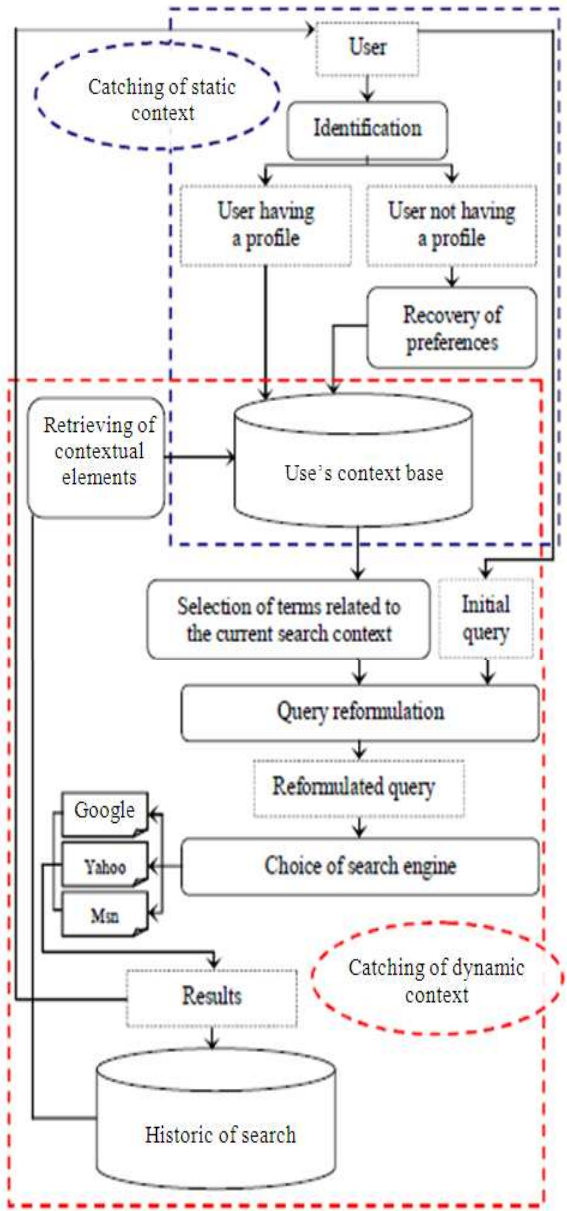

Fig. 1: General architecture of the system for the contextual query reformulation

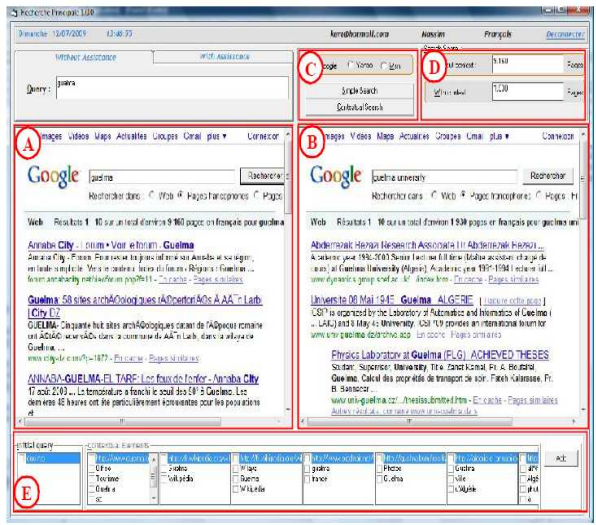

Fig. 2: The main interface of the prototype

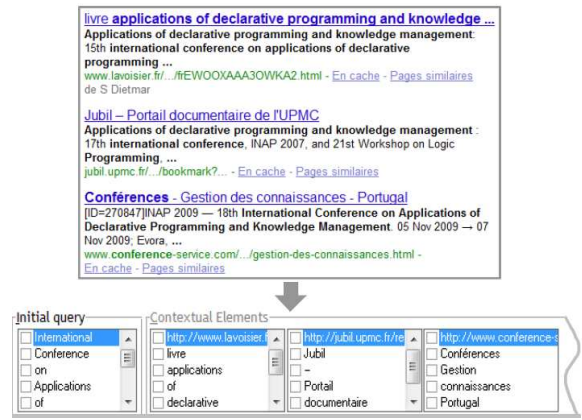

Fig. 3: Principle used to catch the dynamic context

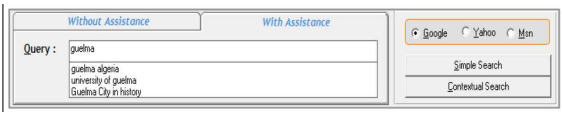

Fig. 4: User's query reformulation

The aim is to compare the scores of relevance in terms of page number returned in the two situations; we note that these scores are calculated by the selected search engine. The interface also offers the possibility to choose the desired search engine (Fig. 2, area C) and gives the score of relevance in terms of number of pages returned in both cases, research with and without and reformulation (Fig. 2, area D). Finally the system captures and offers the contextual elements witches are used to enrich the dynamic context of the user (Fig. 2, area E). Figure 2 presents the main interface of the developed prototype.

**Catching of dynamic context:** To catch the dynamic context, the system automatically analyzes the content of the returned web page and proceeds as follows:

- Retrieves the title of each result and proceeds to its segmentation into a set of words
- It eliminates the empty words using an anti-dictionary
- It proposes the obtained words witches are supposed relevant to enrich the dynamic context to the user
- Finally the user validates terms those he considers truly relevant among all proposals and the system increases the content of the dynamic context

**User's query reformulation:** To ensure the reformulation of the initial user's query, the system retrieves the user's input and compares it with the contents of its context database, after the system retrieves all attribute-value pairs whose left side is identical to the initial query and offers the right side as opportunities of reformulation proper to the current context search. The content of the list of proposals is updated as the user progresses through the seizure of his query. Finally, the best form of reformation is used and the search result is displayed. The Fig. 4 presents the user's query reformulation process.

## DISCUSSION

To carry out the evaluation of our system we performed the same set of 15 queries, both on Google, yahoo and Bing. These queries were made up of 10 simple scenarios covering the range of current needs of a user (they were simple applications of thematic travel, consumption, news and culture) and 5 complex scenarios (rare word or specialized search). Each time, the first ten results displayed have been examined and evaluated for both research, with and without reformulation.

We evaluated the performance of our system according to the relevance of the top 10 results in the three following criteria:

- Criterion 1 (C1): the relevance of the first three results
- Criterion 2 (C2): the relevance of the last seven results

Table 3: The role of the queries reformulation on the three search engines

| | Google | | Yahoo | | Bing | |
|---|---|---|---|---|---|---|
| | Without | With | Without | With | Without | With |
| C1 | 6.62 | 7.69 | 5.78 | 6.11 | 3.38 | 4.23 |
| C2 | 5.60 | 6.77 | 4.92 | 4.18 | 3.94 | 4.87 |
| C3 | 7.40 | 8.19 | 7.56 | 8.55 | 5.54 | 6.22 |

- Criterion 3 (C3): the rate of redundant results (or from the same site)

Each criterion (Cwas rated on 10. Each query received an overall rating out of 30 (3 criteria rated on 10). Each search engine has received-in both cases of research, with and without reformulation-a note on 450 points (30×15 scenarios) reduced to a note on 10. Table 3 shows the results of this evaluation.

Table 3 shows that for the relevance of the first three results with the three search engines, searching with reformulation gives a higher score than that obtained without reformulation. We also note that google tops the list with an improvement of 1.07 point where the query is reformulated.

On the relevance of the last seven results, the two search engines Google and Bing are better adapted to our reformulation system by giving a score higher with the use of a reformulated query, the score increases by 1.17 point in the case of google and by 0.68 in the case of Bing. In contrast the reformulation of query decreases the score of relevance in the case of yahoo by -0.24, this decrease is because the search algorithm used by yahoo works better with a query containing fewer keywords.

Finally, with regard to the rate of redundant results, the three search engines have achieved a higher score after query reformulation. The gap between research without reformulation and with reformulation is respectively 0.79 in the case of google, 0.99 in the case of yahoo and 0.69 in the case of Bing.

## CONCLUSION

The aim of this research is to investigate and discuss the consideration of user's context in information retrieval. In this study we proposed a contextual query reformulation system that catches the user's context according to his profile to modify his initial query. First, we presented the notion of context for information retrieval system. After that, to clarify the concept of user's profile, we defined and classified the user's profiles according to their use in the information retrieval domain. During the presentation of our system, we justified our choice of system parameters, presented our architecture, showed how the static and the dynamic contexts are caught and illustrated the manner in which these two contexts are mutually used in our proposal. Finally, we described a scenario of a contextual reformulation as processed by our system.

This study paves the way for diverse perspectives that are located on two levels: deepening of realized research and enlargement of its application field. With regard to the deepening of the proposed work, it would be interesting to improve the developed prototype in the sense that catching the elements related to the user's dynamic context is done without his intervention and in terms of the application field expansion, it would be interesting to test the proposed architecture on several search engines to measure the contribution of contextual reformulation on each of them and focus the prototype on the search engine that adapts best to our proposal.